\documentclass[sigconf]{acmart}

\AtBeginDocument{%
  \providecommand\BibTeX{{%
    \normalfont B\kern-0.5em{\scshape i\kern-0.25em b}\kern-0.8em\TeX}}}

\copyrightyear{2023}
\acmYear{2023}
\setcopyright{acmlicensed}\acmConference[SIGIR '23]{Proceedings of the 46th International ACM SIGIR Conference on Research and Development in Information Retrieval}{July 23--27, 2023}{Taipei, Taiwan}
\acmBooktitle{Proceedings of the 46th International ACM SIGIR Conference on Research and Development in Information Retrieval (SIGIR '23), July 23--27, 2023, Taipei, Taiwan}
\acmPrice{15.00}
\acmDOI{10.1145/3539618.3591859}
\acmISBN{978-1-4503-9408-6/23/07}

\begin{document}

\title{Delving into E-Commerce Product Retrieval with Vision-Language Pre-training}

\author{Xiaoyang Zheng}
\affiliation{%
  \institution{Alibaba Group}
  \city{Hangzhou}
  \country{China}
    \institution{zhengxiaoyang.zxy@alibaba-inc.com}
}

\author{Fuyu Lv}
\affiliation{%
  \institution{Alibaba Group}
  \city{Hangzhou}
  \country{China}
    \institution{fuyu.lfy@alibaba-inc.com}
}

\author{Zilong Wang}
\affiliation{%
  \institution{Alibaba Group}
  \city{Hangzhou}
  \country{China}
    \institution{huanshi.wzl@alibaba-inc.com}
}

\author{Qingwen Liu}
\affiliation{%
  \institution{Alibaba Group}
  \city{Hangzhou}
  \country{China}
    \institution{xiangsheng.lqw@alibaba-inc.com}
}

\author{Xiaoyi Zeng}
\affiliation{%
  \institution{Alibaba Group}
  \city{Hangzhou}
  \country{China}
    \institution{yuanhan@taobao.com\\
    }
}

\renewcommand{\shortauthors}{Zheng, et al.}

\begin{abstract}
E-commerce search engines comprise a retrieval phase and a ranking phase, where the first one returns a candidate product set given user queries. Recently, vision-language pre-training, combining textual information with visual clues, has been popular in the application of retrieval tasks. In this paper, we propose a novel V+L pre-training method to solve the retrieval problem in Taobao Search. We design a visual pre-training task based on contrastive learning, outperforming common regression-based visual pre-training tasks. In addition, we adopt two negative sampling schemes, tailored for the large-scale retrieval task. Besides, we introduce the details of the online deployment of our proposed method in real-world situations. Extensive offline/online experiments demonstrate the superior performance of our method on the retrieval task. Our proposed method is employed as one retrieval channel of Taobao Search and serves hundreds of millions of users in real time.
\end{abstract}

\ccsdesc[500]{Information systems~Retrieval models and ranking}
\ccsdesc[500]{Information systems~Multimedia and multimodal retrieval}

\keywords{Multimodal Pre-training, Semantic Retrieval, Representation Learning}


\maketitle
\section{Introduction}

Nowadays, online shopping is quite common in our daily lives, and thus creates billions of requests to E-commerce platforms every day, such as Amazon, eBay, Taobao, etc. The search engine for E-commerce products, which displays products given user queries has become the bridge between online sellers and individual buyers. Considering searching efficiency and latency, a searching service consists of two phases: the retrieval phase and the ranking phase. The retrieval phase selects a candidate set from all documents, and then the ranking phase determines the displaying orders. The efficiency of the retrieval system imposes an upper bound on the overall performance of the search engine. In Taobao (the top E-commerce platform in China), the whole documents contain billions of products. Each one has one title in Chinese and several displaying images, while user queries are plain texts. In this paper, we focus on the retrieval phase of Taobao Search and attempt to solve the text-to-multimodal matching problem.

In recent years, the paradigm of pre-training and fine-tuning has shown great potential in the area of information retrieval \cite{yu2022commercemm,gao2020fashionbert,zheng2023make,DBLP:conf/kdd/ZouZCMCWSCY21}. Especially the power of transformer \cite{vaswani2017attention} and vision-language representation learning has motivated people to study the application of pre-training on E-commerce product search \cite{gao2020fashionbert,chia2022contrastive,yu2022commercemm}. However, those previous works directly apply pre-training techniques, 
but rarely consider their scalability on large-scale problems. 
For instance, an academic retrieval task requires dozens of candidates from thousands of documents, while in Taobao Search, the retrieval task returns tens of thousands of candidates from billions of documents. 
Therefore, in order to deal with the large-scale retrieval problem in real-world applications, we introduce two negative sampling strategies: Memory Bank Negative Sampling and Cross-Device Negative Sampling.
Different from the common in-batch negative sampling~\cite{DBLP:conf/icml/ChenK0H20}, relying on the size of mini-batches and device memory, negative sampling schemes adopted in this paper not only facilitate the ability of retrieval in a large corpus but also help produce stable features.
Moreover, previous pre-training methods~\cite{zheng2023make,gao2020fashionbert} for E-commerce retrieval apply regression-based image reconstruction tasks. However, unlike generative tasks such as image manipulation~\cite{Patashnik_2021_ICCV} and text-to-image generation~\cite{DBLP:journals/corr/abs-2204-06125}, the retrieval task requires the ability to distinguish the difference within the visual feature space.
Therefore, we propose a novel visual pre-training task based on contrastive learning and thus boost the representations of images.


To this end, we propose a vision-language pre-training model with a novel visual pre-training task and negative sampling schemes to solve the text-to-multimodal matching problem in E-commerce search engines. 
Besides, in this paper, we introduce the system architecture of Taobao Search and the online deployment of our proposed model. Extensive offline/online experiments demonstrate the efficiency of our method.


\section{The Proposed Method}
\begin{figure}[htbp]
    \centering
    \includegraphics[width=0.38\textwidth]{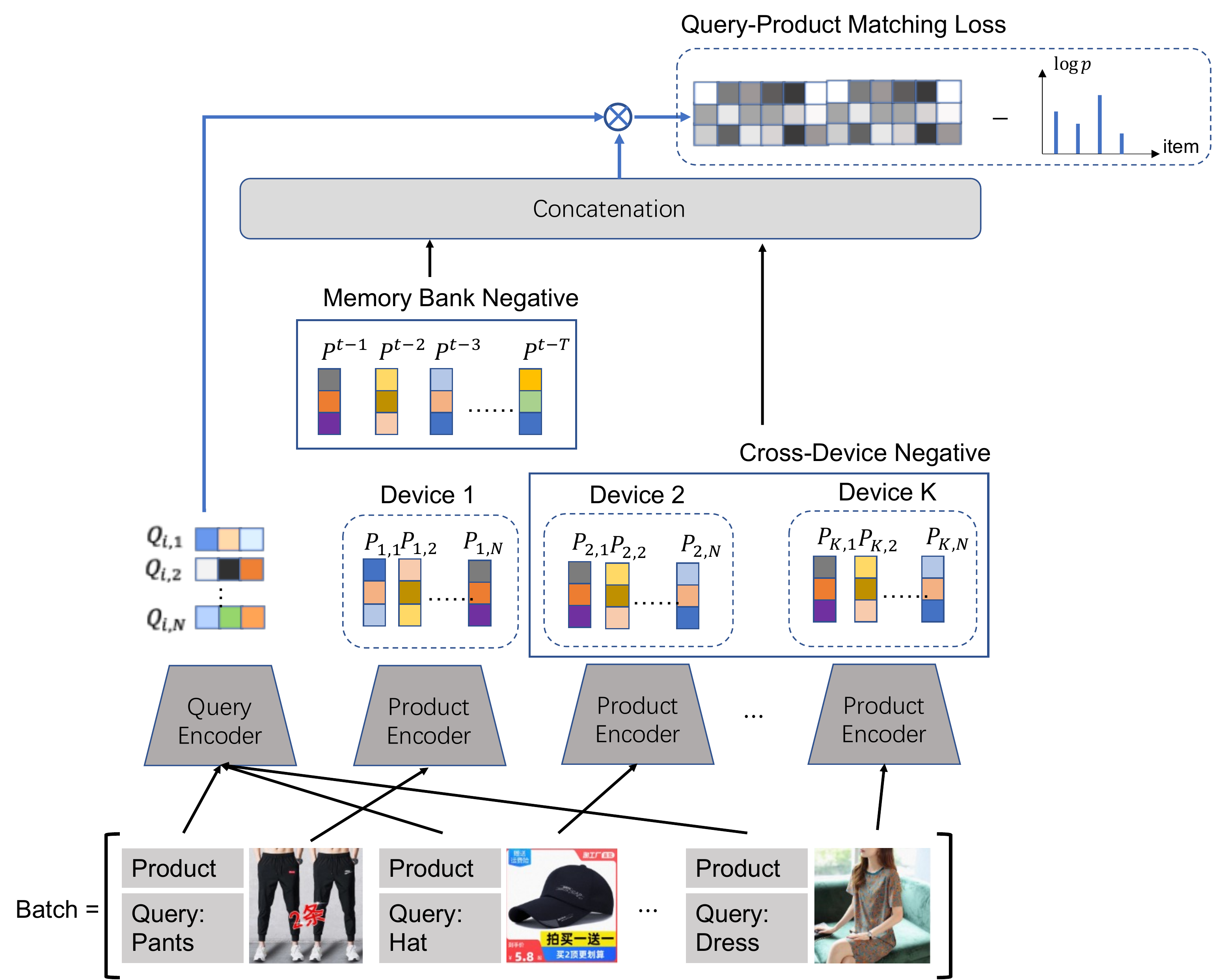}    
    \caption{The architecture of our proposed method with CDNS and MBNS.}
    \label{fig:network}
\end{figure}
\subsection{Model Overview}

\paragraph{\textbf{Model Inputs. }} Taobao Search mainly deals with Chinese information. The textual inputs (user queries and product titles) to our model are firstly segmented into Chinese words and sorted in lexicographical order. These words are tokenized into text sequences (\textbf{Q} for user queries and \textbf{T} for product titles) following the way of BERT \cite{devlin2019bert}. We adopt the Chinese vocabulary provided in \cite{qiu2021easytransfer}. The product images are divided into 4x4 patches and then extracted by a ResNet-101 backbone model into image sequences (\textbf{I} for product images). All textual and visual sequences are converted into embeddings, respectively.

\paragraph{\textbf{Model Architecture. }} Our retrieval model consists of separated encoders for user queries, product titles, and product images, respectively, as shown in Figure \ref{fig:network}. The title encoder and the image encoder are presented together as the product encoder concisely, composed of 12 layers of transformers \cite{vaswani2017attention}.
The outputs from [CLS] tokens are regarded as representations of the whole sequence.

\begin{figure}[htbp]
    \centering
    \includegraphics[width=0.3\textwidth]{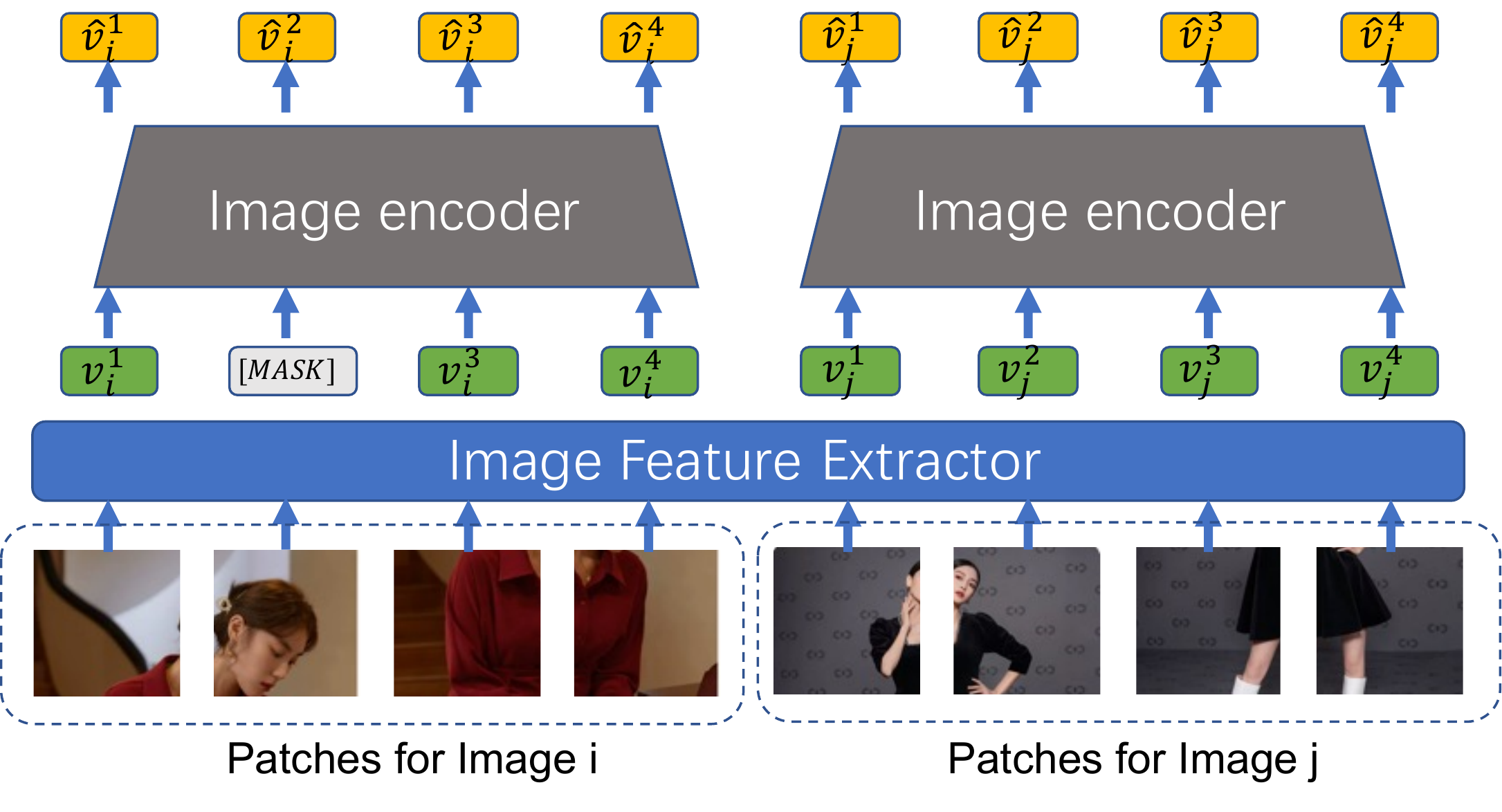}
    \caption{The visual pre-training task based on contrastive learning.}
    \label{fig:image}
\end{figure}
\paragraph{\textbf{Pre-training Tasks. }} Following MAKE \cite{zheng2023make}, we apply two self-supervised pre-training tasks (Masked Language Modeling and Masked Patch Modeling) and two supervised pre-training tasks (Query-Product Matching and Query-Product Classification). Please refer to the paper for detailed information. However, the previous MPM task reconstructs image embeddings based on only the context from the same image and thus lacks the ability to differentiate from other images. To this end, we design a novel visual pre-training task based on contrastive learning, which jointly optimizes the model upon various visual information. Given input image embedding sequence with a random mask $\mathbf{v}=\{v_i^1,v_i^2,\cdots,[\text{MASK}],v_i^L\},i\in[1,N]$, the transformer returns a sequence of predicted image embeddings $\mathbf{\hat{v}}$ correspondingly, as shown in Figure \ref{fig:image}. $N$ is the batch size and $L$ is the length of image sequences. The Masked Patching Modeling (MPM) loss adapted to the retrieval task ($\text{MPM}_R$) is presented as:
\begin{align}
\tiny
    p(v,\hat{v})&=\log\frac{\exp(v\cdot\hat{v}/\tau)}{\sum_{i=1}^N\sum_{j=1}^{L}\exp(v\cdot\hat{v}_i^j/\tau)}, \\
    L_{MPM_{R}}&=-\frac{1}{N_{\text{MASK}}}\sum p(v_i,\hat{v}_i),
\end{align}
where $\tau$ is a trainable temperature parameter and $N_{\text{MASK}}$ is the number of [MASK] tokens in one batch. With the new image pre-training task, our model produces suitable image embeddings adapted to the retrieval task, compared against regression-based pre-training task \cite{gao2020fashionbert}.

\subsection{Negative Sampling Schemes}
It is widely studied that the size of negative samples is the key to contrastive learning \cite{DBLP:conf/icml/ChenK0H20}. Following MAKE \cite{zheng2023make}, we adopt the query-product matching loss, benefiting from contrastive learning. However, in-batch negative sampling adopted in MAKE \cite{zheng2023make} chooses positive samples out of dozens of negative samples, and thus is unsuitable for a large-scale retrieval task. Therefore, we employ two schemes to enlarge negative samples, Cross-Device Negative Sampling (CDNS) and Memory Bank Negative Sampling (MBNS) to boost representation learning.

\paragraph{\textbf{Cross-Device Negative Sampling.}} As studied in \cite{DBLP:conf/sigir/WangZH21}, the embeddings become relatively stable after some iterations and thus provide valid information for training. As a result, we gather the product embeddings from all devices during distributed training. For one positive query-product pair, the product embeddings from other devices are regarded as negative samples.
In our implementation, we take advantage of a built-in interface of Whale Estimator \cite{DBLP:conf/usenix/JiaJWXS0LCLZL022}, a powerful framework for distributed training. We only apply Cross-Device Negative Sampling after $50\%$ of all training iterations for stable features.

\paragraph{\textbf{Memory Bank Negative Sampling.}} Similar to MoCo \cite{DBLP:conf/cvpr/He0WXG20}, strong representations of queries and products are learned by extracting rich information from immediate preceding batches. In our implementation, we maintain a first-in-first-out queue to conduct MBNS. At the end of each iteration, the current mini-batch is pushed into the queue and the oldest mini-batch is removed. The whole queue is regarded as a set of negative samples. 
Different from \cite{DBLP:conf/cvpr/He0WXG20}, we do not adopt the momentum update, and negative samples are chosen from the queue without back-propagation.
For the consistency of representations, we only apply MBNS after $60\%$ of total training steps and avoid negative samples from an unstable encoder. Although the encoder is not updated instantly with MBNS samples, the model still benefits from large MBNS samples.  

Compared with the common in-batch negative sampling, CDNS and MBNS enlarge the number of negative samples (from dozens to tens of thousands) during training, and thus lead to stronger ability in discriminative tasks, such as E-commerce retrieval tasks studied in this paper. The softmax output for a query-product pair $(u,v)$ is modified as: 
\begin{equation}
\small
    p(u,v)=\log\frac{e^{s(u,v)}}{e^{s(u,v)}+w_1\sum_{i=1}^{N_1}e^{s(u,v_i)}+w_2\sum_{j=1}^{N_2}e^{s(u,v_j)}},
\end{equation}
where $s(u,v)=u^Tv-\log p$ is an extended inner-product similarity between query/product embeddings. The $\log p$ term represents the sampling probability of the product, with which our model avoids excessive attention on popular products. Due to the convergence over time, CDNS provides more informative negative samples than MBNS. Therefore, we assert different weights for these two negative sampling schemes. ($w_1=0.25$ and $w_2=1$ in this paper). $N_1$ and $N_2$ are the numbers of negative samples from MBNS and CDNS, respectively.

\section{Online Serving}
\begin{figure}[htbp]
    \centering
    \includegraphics[width=0.38\textwidth]{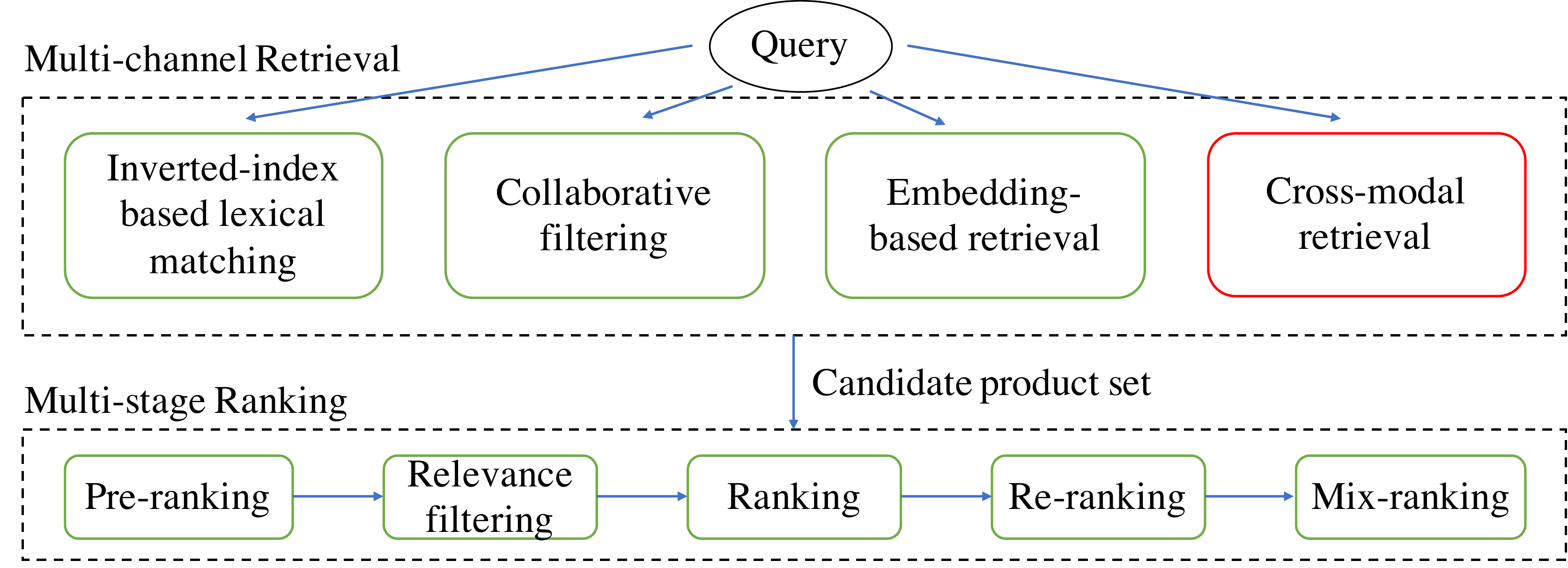}
    \caption{System architecture of our searching system.}
    \label{fig:online}
\end{figure}

As shown in Figure \ref{fig:online}, our search system works in a multi-stage pipeline. When a user issues a query, the multi-channel retrieval system returns an unordered candidate product set. Previously, the retrieval system contains three channels: inverted-index-based lexical matching, collaborative filtering, and embedding-based deep semantic retrieval. The retrieval system merges and de-duplicates results from three channels, and then passes candidates to the ranking stage, including pre-ranking, relevance filtering, ranking, re-ranking, and mix-ranking. Different from the existing three-channel retrieval, our pre-training method considers both semantic and visual information of products. Our proposed method serves as the fourth channel in the retrieval system. 

\subsection{Offline Prediction and Indexing}
We follow an offline-to-online pipeline to deploy the model. At the offline stage, all product embeddings are exported from the product encoder and then passed to Proxima, an ANN (approximate nearest neighbor) framework.
%
%
The ANN framework builds indexes of product embeddings with HC (hierarchical clustering) algorithm. We collect queries from online clicking logs for one month and feed them to the query encoder. With offline indexes, query embeddings are used to retrieve the most related top-K products. 
The model parameters and the offline indexes are updated weekly.


All queries from the offline query-product pairs are segmented into term sequences and then sorted in lexicographical order. The term sequences are hashed into 32-bit integers as query ids (qid) and attached to related products. With the pre-processing of query segmentation and sorting, our model is robust to queries with misspellings, such as \textit{dress red} vs. \textit{red dress}. The offline indexes are published to the search engine service HA3 regularly. 

Once the retrieval system is triggered by a query, the backend Query Processing service transforms the query into a qid, and HA3 retrieves products with the published index. One advantage of our deployment is that the inference time and computational cost are negligible.

\begin{figure}[htbp]
    \centering
    \includegraphics[width=0.38\textwidth]{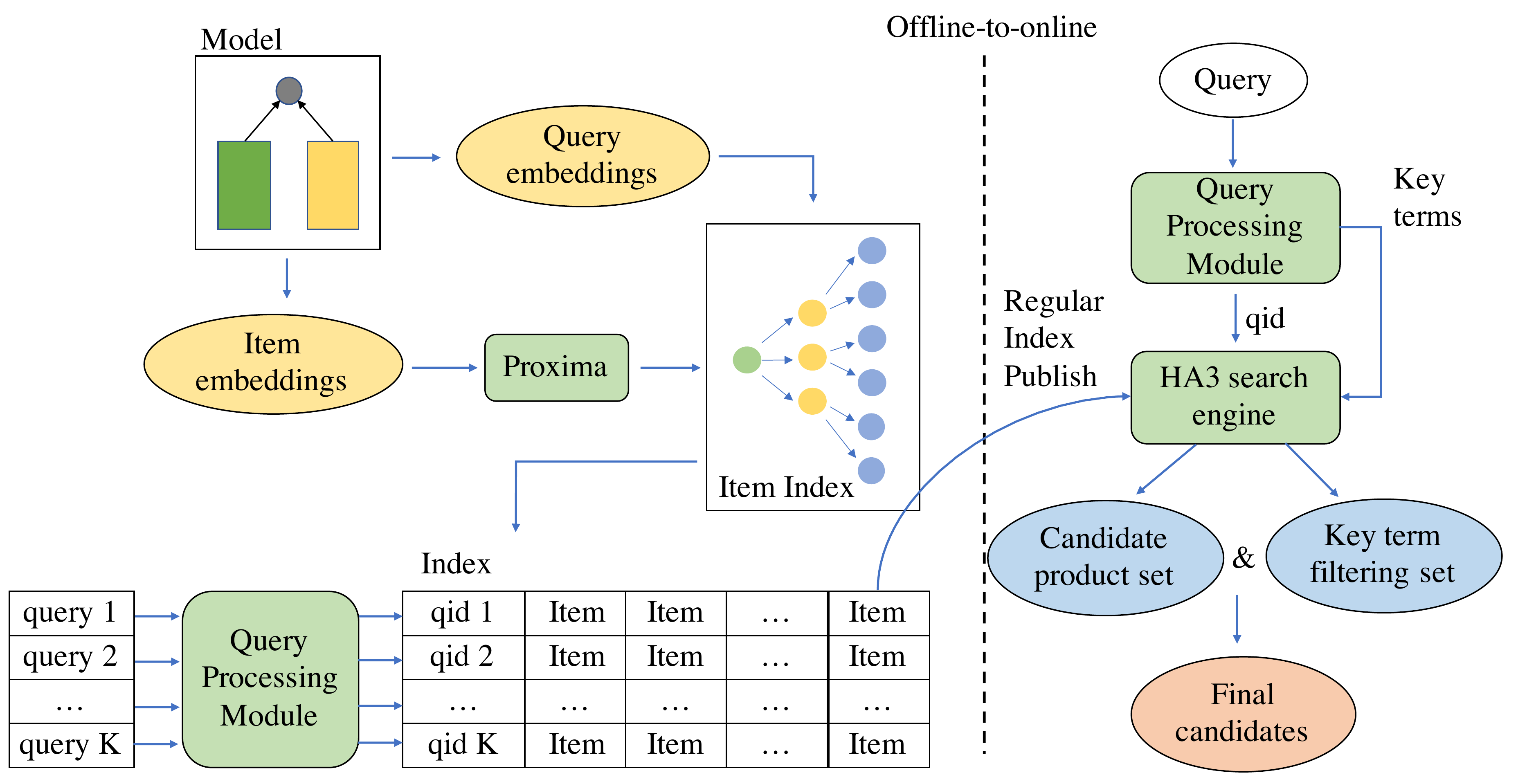}
    \caption{Deployment system of our proposed method.}
    \label{fig:serve}
\end{figure}

\subsection{Relevance Control}
In practical usage, we need to further consider queries with exclusive terms, such as brands. For instance, given a query \textit{Nike Shoes}, the candidates of \textit{Adidas Shoes} are strictly forbidden from the retrieval results. It is widely known that retrieval methods with dual encoders and ANN indexing fail to handle these bad cases. To this end, a Boolean matching module based on inverted indexes is adopted to further filter ANN results. The constraints of key terms are applied by using logical AND operations. For example, search results of a query \textit{Nike shoes} are expressed as:
\begin{text}(ANN results) AND (Brand: Nike) AND (Category: Shoes)\end{text}. The restricted terms are extracted according to query understanding and inserted into the query language of HA3. This strategy of relevance controlling significantly reduces the number of bad cases and improves user experience.

\section{Experiment}
\subsection{Datasets, Implementations, and Metrics}
\paragraph{\textbf{Dataset. }} We collect online clicking and purchasing logs with user queries, product titles, and images from Taobao Search. In order to prevent the Matthew Effect, query-product pairs are randomly chosen on long-tail samples as the training set, which is indeed a user-annotated dataset with relevant query-product pairs. One million samples are randomly sampled as the evaluation set. Besides, in order to measure the model generalization in other scenarios, we sample one million clicked products from the recommendation scenario from Taobao App, where each product is associated with a mocked query. 

\paragraph{\textbf{Model Implementation. }} Our proposed model is composed of 12 layers of transformers \cite{devlin2019bert}, where each layer has $768$ hidden units and $12$ self-attention heads. We start training from a StructBERT \cite{wang2019structbert} checkpoint, pre-trained on Chinese E-commerce corpus. The model is trained for 5 epochs on 40 NVIDIA V100 GPUs. 
We employ an AdamW \cite{DBLP:journals/corr/abs-1711-05101} optimizer with $\beta_1=0.9$ and $\beta_2=0.98$. The learning rate is warmed-up to $1e^{-4}$ in the first $20\%$ iteration and decays to $0$, both following a linear schedule.

\paragraph{\textbf{Offline Evaluation Metrics. }} Given a user query $q_i$, the top-K retrieved product set is denoted as $R_{i,K}=\{p_1,p_2,\ldots,p_K\}$. The clicked/purchased products from the evaluation set are denoted as the target set $T_i$. A Recall@K metric is applied to measure the retrieval performance:
\begin{equation}
\small
    \text{Recall@K}=\frac{1}{N}\sum_{i=1}^{N}\mathbb{I}(\exists t | t\in R_{i,K} \wedge t\in T_i),
\end{equation}
where $\mathbb{I}(\cdot)$ is an indicator function and $N$ is the size of the evaluation dataset. 

We also use two relevance metrics $P_{rel}$ and $P_{cate}$, measuring the overall quality of the retrieved products. The first one is computed according to a well-trained relevance model \cite{DBLP:conf/www/YaoTCYXD021}. 
\begin{equation}
\small
    P_{rel}=\frac{1}{NN_R}\sum_{i=1}^N\sum_{j=1}^{N_R} f(q_i,p_{i,j}),
\end{equation}
where $f(\cdot,\cdot)\in [0,1]$ returns the predicted probability from the relevance model. $N_R$ is the size of the retrieval set $R$. 
For the second metric, the intent categories based on user queries are predicted according to statistical behaviors. The second metric measures the rate of consistency between the intent categories and the categories of the retrieved products. 
\begin{equation}
\small
    P_{cate}=\frac{1}{NN_R}\sum_{i=1}^N\sum_{j=1}^{N_R} \mathbb{I}(f_c(q_i)=c_{i,j}),
\end{equation}
where $f_c(\cdot)$ denotes the intent categories based on user queries, and $c_{i,j}$ is the category of the retrieved product $p_{i,j}$

\paragraph{\textbf{Online Evaluation Metrics. }} The number of transactions (denoted as \#Trans) and GMV (Gross Merchandise Volume, total value of sales) are employed as online evaluation metrics. 
Besides, we apply $P_{rel}$ computed on online retrieval candidates to measure relevance. We also use additional latency and memory to assess the cost of our method.

\subsection{Offline Experimental Results}
\paragraph{\textbf{Comparison with Baseline Methods. }} We choose FashionBERT \cite{gao2020fashionbert}, CLIP \cite{radford2021learning}, CommerceMM \cite{yu2022commercemm} and MAKE \cite{zheng2023make} as baseline methods.
The first three baseline methods on E-commerce retrieval tasks represent pre-training architectures of uni-encoder, dual-encoder, and mixed-encoder, respectively. Instead of FashionCLIP \cite{chia2022contrastive}, we adopt CLIP since FahionCLIP focuses on the fashion industry, while we aim at evaluating the performance on the general E-commerce domain. 
All methods are fine-tuned on the same dataset for fair comparisons. As shown in Table \ref{tab:ablation}, our proposed method, designed for the text-to-multimodal task in real-world situations, outperforms all baseline methods in terms of retrieval efficiency and relevance.

\paragraph{\textbf{Ablation Study. }} We apply MAKE \cite{zheng2023make} as the baseline in ablation study. The novel visual pre-training task $\text{MPM}_R$ enforces the model to distinguish the difference across images and thus boost the retrieval performance ($+0.32\%$ in relevance and $+1.22\%$ in recall hitrate). In addition, compared with MAKE \cite{zheng2023make}, both CDNS and MBNS facilitate the recall hitrate on the evaluation set. With the greatly increased number of negative samples, the model is able to retrieve more accurate products. However, we observe few improvements in the recall relevance. Since these negative samples are gathered across different devices and iterations, the convergence gap of encoders compromises the quality of products retrieved at the tail. The combination of all optimization techniques improves the recall hitrate by $\textbf{+0.33\%}$ and the recall efficiency by $\textbf{+2.61\%}$.

\begin{table}[htbp]
    \centering
    \footnotesize
    \begin{tabular}{lccc}
    \hline
        Methods & $P_{rel}\uparrow$ & $P_{cate}\uparrow$ & Recall@$K\uparrow$ \\\hline
        FashionBERT \cite{gao2020fashionbert} & 0.8385 & 0.8190 & 0.3867 \\
        CLIP \cite{radford2021learning} & 0.8648 & 0.8423 & 0.4675\\
        CommerceMM \cite{yu2022commercemm} & 0.8710 & 0.8653 & 0.4937 \\
        MAKE \cite{zheng2023make} & 0.9014 & 0.9295 & 0.6088 \\\hline
        CDNS & 0.9018 & 0.9272 & 0.6318 \\
        MBNS & 0.9010 & 0.9273 & 0.6171 \\
        $\text{MPM}_{R}$ & 0.9046 & 0.9319 & 0.6210 \\\hline
        Ours & \textbf{0.9047} & \textbf{0.9324} & \textbf{0.6349} \\\hline
    \end{tabular}
    \caption{Offline experimental results and ablation studies.
    }
    \label{tab:ablation}
\end{table}

\subsection{Online A/B Tests}
We deploy our proposed model as one retrieval channel of Taobao Search. As shown in Table \ref{tab:AB}, our proposed method improves the retrieval system on GMV and \#Trans by $+0.46\%$ and $+0.39\%$, respectively. Besides, our method boosts the overall relevance of online retrieval candidates by $+1.20\%$. Our method introduces negligible additional cost (2 ms in latency and $12$ GB in memory cost) compared to the gains in efficiency. All these online A/B results are averaged over more than one month. These results demonstrate that our proposed method significantly improves the overall performance of Taobao Search.
\begin{table}[htbp]
    \centering
    \footnotesize
    \begin{tabular}{l|ccccc}
    \hline
        Methods & GMV & \#Trans & $P_{rel}$ & Time Cost & Memory Cost \\\hline
        Ours & +0.46\% & +0.39\% & +1.20\% & +2 ms & +12 GB\\\hline 
    \end{tabular}
    \caption{Online A/B tests on Taobao Search. 
        }
    \label{tab:AB}
\end{table}



\section{Conclusions}
In this paper, we propose a novel vision-language pre-training method tailored for the text-to-multimodal retrieval task of Taobao Search. We employ a novel visual pre-training task specially designed for the retrieval problem, which underlines the difference within the visual feature space instead of the ability of reconstruction in generative models. In addition, we apply two different negative sampling schemes (cross-device negative sampling and memory bank negative sampling), to distinguish relevant products from a large corpus. Besides, we introduce the online architecture and our practice of deployment. Offline experiment results and online A/B tests demonstrate the great performance of our proposed method. 

\section*{COMPANY PORTRAIT}
Alibaba Group aims to enable small enterprises to leverage innovation and technology to grow and compete more effectively in the domestic and global economies. Since launching its first website helping small and medium-sized enterprises in China to sell internationally, Alibaba Group has grown into a digital ecosystem with businesses comprising China commerce, international commerce, local consumer services, Cainiao, cloud, digital media, entertainment, innovation initiatives, and others.

\section*{BIOGRAPHY}
Xiaoyang Zheng is an algorithm engineer at Alibaba Group, where he works on multimodal retrieval in Taobao Search. Xiaoyang Zheng received a master's degree and a bachelor's degree in computer science from Shanghai Jiao Tong University. His research interests include information retrieval and multimodal representation learning.

\bibliographystyle{ACM-Reference-Format}
\balance
\bibliography{sample-base}

\end{document}